\documentstyle[sprocl]{article}

\input{psfig}

\def\be{\begin{equation}}
\def\ee{\end{equation}}
\def\bea{\begin{eqnarray}}
\def\eea{\end{eqnarray}}

\begin{document}

\title{ BUBBLE NUCLEI, NEUTRON STARS AND QUANTUM BILLIARDS 
\footnote{Talk at {\it Bologna 2000 --
Structure of the Nucleus at the Dawn of the Century},
     May 29 -- June 3, 2000, Bologna, Italy} }

\author{Aurel BULGAC }

\address{Department of Physics, University of Washington, Seattle WA
98195--1560, USA  \\ E-mail: bulgac@phys.washington.edu } 
\address{Max--Planck--Institut f\"ur Kernphysik, Postfach 10 39
          80, 69029 Heidelberg, GERMANY }

\author{Piotr MAGIERSKI }

\address{Institute of Physics, Warsaw University of Technology,
          ul. Koszykowa 75, PL--00662, Warsaw, POLAND\\
          E--mail: magiersk@if.pw.edu.pl }

\maketitle\abstracts{ We briefly review the significance of quantum
corrections in the total energy of systems with voids: bubble nuclei,
atomic clusters and the inhomogeneous phase of neutron stars.}

It was suggested a long time ago that very large nuclei might not
undergo a Coulomb explosion if they acquire a new topology, that of a
bubble or a torus \cite{wheeler}. When a void is formed, while the
density and therefore the total volume is kept unchanged, the surface
area of such a nucleus naturally increases and that leads to an
increased surface energy and less binding. However, at the same time
the average distance between protons increases as well and the total
Coulomb energy then decreases. The balancing of these two types of
energy and the fact that configurations with larger binding energy
than the familiar compact geometries exists is the reason why
bubble--like and torus--like nuclei could in principle be someday
observed.  It was realized however that shell effects play a crucial
role in stabilizing these new shapes \cite{wong}. During the last
decade many experimentalists have tried to manufacture highly charged
metallic clusters, but, again, Coulomb repulsion prevented their
creation. The idea that objects with a different topology, in
particular bubble--like charged metallic clusters could be a possible
route to create highly charged metallic clusters was recently put
forward \cite{dietrich}, and again the stabilizing role of the shell
corrections was noted as playing a decisive role.

There was an aspect of bubble systems, which for mysterious reasons
never caught the attention of previous authors:{\em Where should one
position a bubble inside a nucleus?} Symmetry considerations seem to
suggest that a spherical bubble should be placed at the center of a
spherical system. A closer look will show however that there is
something more than mere symmetry and that Coulomb energy plays
perhaps the most important role in stabilizing the bubble position. It
is relatively straight forward to show that if one were to displace a
bubble from the center of a nucleus the Coulomb energy would
increase. When considering Coulomb effects, one can think of a bubble
as being a charged object, having the same charge density as the rest
of the matter, but of opposite sign. One can then easily evaluate the
Coulomb force acting on a bubble. Inside a spherical uniformly charged
object the electric field is radial and can be easily be evaluated
using Gauss law:
\be
{\bf{E}}({\bf{r}}) = \frac{4 \pi \rho _0{\bf{r}}}{3},
\ee
where $\rho _0$ is the charge density.  Thus the force acting on a
bubble is simply the integral over the bubble ``effective charge''
times the electric field
\be
{\bf{F}}_b = -\int_{bubble}  d^3r \frac{4 \pi \rho _0^2{\bf{r}}}{3}=
-\frac{4\pi \rho _0^2 V_b{\bf{R}}}{3}, 
\ee
where $V_b$ is the volume of the bubble and ${\bf{R}}$ is the position
vector of the bubble center, with respect to the nuclear center.  When
the suggestion was made to make cavities inside charged metallic
clusters it became clear to us that the above argument is incomplete
\cite{bub1,bub2} and there is no apparent physical candidate
responsible for determining the optimal bubble position inside a
homogeneous fermi system.  As freshmen physics students know, there is
no electric field inside a metal in the absence of electric
currents. If in the case of nuclei one could invoke, either symmetry
arguments (for some not totally clear reasons) or, better yet, the
stabilizing role of Coulomb force, it was not obvious what made a
bubble system stable in the case of a metal cluster. None of the
``usual suspects'' (volume, surface, curvature or Coulomb energies)
seem to play any significant role and one might naturally expect that
if there is something happening in a metal cluster, a similar
mechanism should most likely be operative in a nucleus as well.  (In
metal clusters one has of course to deal with additional ionic degrees
of freedom, however, many cluster properties are determined mostly by
the electrons alone and the ions are merely spectators.)  The solution
to the above puzzle was rather simple, but at the same time to a large
extent unexpected as well: the physics of a bubble is governed by pure
quantum effects, known in nuclear physics as shell corrections and in
quantum filed theory as Casimir energy \cite{casimir}. Instead of
presenting formulas and results of numerical calculations we shall
limit ourselves here to a general discussion of some of the novel
aspects of these systems and refer the interested readers to the
available references.

When one mentally starts pushing a bubble around inside a finite fermi
system one obviously excites such a system, if initially the bubble
was in its optimal position and therefore the entire system in its
ground state. Since the displacement of a bubble will affect many
particles, bubble displacements are naturally collective
excitations. Apart from collective pairing excitations, perhaps no
other collective mode in a fermi system is purely quantum in
nature. Since shell corrections effects scale with particle number as
$\propto N^{1/6}$, see Ref. \cite{strutin}, and other collective modes
involve some degree of surface deformation, and therefore their
effects scale with particle number as $\propto N^{2/3}$, one can
expect that bubble displacements would correspond to perhaps the
softest collective modes possible. Our vast experience seem at this
point to lend support to the idea that symmetry should play a major
role in determining the optimal position of a bubble, since shell
correction effects are largest for spherical systems. To some extent
this is true, see Ref. \cite{bub2}, however with many provisos. Even
if a system is ``magic'', once one would displace a bubble off center
significantly, the potential energy surface becomes rather flat. One
would also expect that the amplitude of the shell corrections will
become smaller when the bubble is significantly off center, since
classically the motion of a particle in the corresponding
single--particle potential is chaotic to a large degree \cite{bohigas}
and the single--particle spectrum is expected to have no large gaps.
As our detailed numerical results show this expectation is hardly ever
true. In all our numerical analyses so far we have used hard wall
potentials (which thus partially explains the origin of the term
quantum billiards in our title), for which there is significant
evidence that they do reproduce the realistic spectra with sufficient
accuracy for the purpose of computing the gross shell structure
\cite{brack}.  A particular feature of the shell correction energy
evaluated for hard wall potential, and which we do not expect to
survive entirely in a selfconsistent calculation, is particularly
interesting however, as it underlines a general trend. We have
observed in Ref. \cite{bub2}, and later confirmed as a general feature
in Refs. \cite{piotr}, that the amplitude of the shell correction
energy increases as the bubble approaches the boundary of the
system. This is particularly puzzling, since the closer the bubble is
to the system boundary the classical motion is more chaotic and one
would naturally expect then the shell energy to decrease, but not to
increase. Part of the explanation is that a particular periodic orbit
becomes prominent and leads to a significant ``scarring'' of the
single--particle density of states. This is the orbit bouncing between
the points of closest approach. The relative size of the bubble also
plays a major role. If the fractional volume of the bubble is small,
then the shell correction energy oscillates with a relatively small
amplitude when compared with a bubble with a larger fractional volume.

Sidestepping the question of bubble stability and of the energy cost
of bubble formation, one can reasonably ask a number of quite relevant
questions as well: ``Why not have a system with two or more bubbles?''
Neutron stars have been predicted a long time ago to have a locally
inhomogeneous phase, often referred to as ``the pasta phase''
\cite{pethick}.  Due to the same type of interplay between the surface
and Coulomb energies, at depths of about 0.5 km below the surface of a
neutron star and at densities just below nuclear saturation density a
new phase is favored, where spherical and rod--like nuclei embedded in
a neutron gas, plates, cylinders and bubbles exist. Almost all
previous analyses of this phase have been performed in the liquid drop
or Thomas--Fermi approximations. It was determined that on the way
inside a neutron star, while the average density is increasing, there
is a well defined sequence of phases: nuclei $\rightarrow$ rods
$\rightarrow$ plates $\rightarrow$ tubes $\rightarrow$ bubbles
$\rightarrow$ uniform matter.  The energy of each of these phases is
significantly below the energy of the uniform phase at the same
average density, irrespective of nuclear model used
\cite{pethick}. The energy differences between various phases even
though are very small, of the order of keV's per fm$^3$, are
apparently independent of the model for the nuclear forces used. The
various models for nuclear forces can lead to significant variations
in the values of the interface surface tension. Shell correction
energy on the other hand is known to be of geometric origin
essentially. Since in infinite matter the presence of various
inhomogeneities does not lead to the formation of discrete levels, one
might call the corresponding energy correction for neutron matter the
Casimir energy \cite{casimir}. The inhomogeneous phase of a neutron
star is basically nothing else but a Sinai billiard, a model which is
widely popular in classical and quantum chaos studies.  In a first
approximation one can treat various objects in the inhomogeneous phase
as spherical, cylindrical or plate like voids in a neutron gas.

In order to better appreciate the nature of the problem we are
addressing here, let us consider the following situation. Let us
imagine that two spherical identical bubbles have been formed in an
otherwise homogeneous neutron matter. For the sake of simplicity, we
shall assume that the bubbles are completely hollow.  We shall ignore
for the time being the role of the Coulomb interactions, as their main
contribution is to the smooth, liquid drop or Thomas--Fermi part of
the total energy.  Then one can ask the following apparently innocuous
question: ``What determines the most energetically favorable
arrangement of the two bubbles?'' According to a liquid drop model
approach (completely neglecting for the moment the possible
stabilizing role of the Coulomb forces) the energy of the system
should be insensitive to the relative positioning of the two
bubbles. Using Gutzwiller trace formula one can show that pure quantum
effects lead to an approximate interaction energy of the following
form
\be
E_{int} = \frac{\hbar^2k_F^2}{2m} \frac{R^2}{\pi a(a-2R)}
     \left \{
  \frac{\cos [2k_F(a-2R)]}{2k_F(a-2R)}
 -\frac{\sin[2k_F(a-2R)]}{4k_F^2(a-2R)^2},
     \right \} \label{eq:eint}
\ee
where $R$ is the bubble radius, $a$ is the distance between the bubble
centers and $k_F$ is the Fermi wave length, see Fig. 1.  It came as
surprise to us to find that two bubbles have a long range
interaction. In condensed matter physics a similar type of interaction
is known for about a half of a century, the interaction between two
impurities in a fermi gas \cite{rk}.  The fact that this interaction
oscillates suggest the intriguing possibility of forming di--bubble
molecules with various radii. However, until one will determine the
inertia of a di--bubble system it is not obvious whether such a
molecule could indeed exist. It can be shown that in the case of three
or more bubbles the interaction among them contains besides the
expected pair--wise interaction we have just described, also genuine
three--body, four--body and so forth interactions. The interaction
Rel. (\ref{eq:eint}) has its origin in the existence of the periodic
orbit bouncing between the two bubbles. In the case of three or more
bubbles there are distinct periodic orbits bouncing between three or
more objects, which are the reason these genuine three and more
body interactions arise.

\begin{center}
\begin{figure}[t]
\psfig{figure=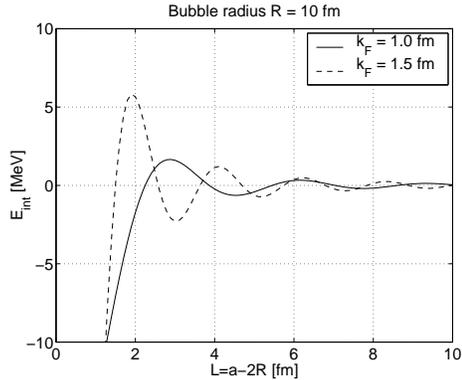,height=2.0in}
\caption{The interaction energy between two bubbles as a function of
the distance between their ``tips''.  \label{fig:eint}}
\end{figure}
\end{center}

Using semiclassical methods (Gutzwiller trace formula), we have
analyzed the structure of the shell energy as a function of the
density, filling factor, lattice distortions and temperature
\cite{piotr}. The main lesson we have learned is that the amplitude of
the shell energy effects is comparable with the energy differences
between various phases determined in simpler liquid drop type models.
Our results suggest that the inhomogeneous phase has perhaps an
extremely complicated structure, maybe even completely disordered,
with several types of shapes present at the same time.

At higher densities in neutron stars one expects that quarks and
mesons will lead to similar structured mixed phases
\cite{heiselberg,glen}. The formation of either quark--gluon droplets
embedded in a hadron gas or of hadron droplets embedded in a
quark--gluon plasma has been studied and predicted for almost a
decade. One naturally expects that similar quantum corrections are
relevant in these cases as well.


\section*{References}

\begin{thebibliography}{99}

\bibitem{wheeler} H.A. Wilson, {\em Phys. Rev.} {\bf 69}, 538 (1946);
J.A. Wheeler, unpublished notes; P.J. Siemens and H.A. Bethe, {\em
Phys. Rev. Lett.} {\bf 18}, 704 (1967); W.J. Swiatecki,{\em Physica
Scripta } {\bf 28}, 349 (1983); W.D. Myers and W.J. Swiatecki, {\em
Nucl. Phys.} {\bf A 601}, 141 (1996).

\bibitem{wong} C.Y. Wong, {\em Ann. Phys.}  {\bf 77}, 279 (1973).

\bibitem{dietrich} K. Pomorski and K. Dietrich, {\em
Eur. Journ. Phys.} {\bf D 4}, 353 (1998).

\bibitem{bub1} A. Bulgac {\it et al.}, in {\it Proc. Intern. Work. on
Collective excitations in Fermi and Bose systems}, eds. C.A. Bertulani
and M.S. Hussein (World Scientific, Singapore 1999), pp 44--61 and
nucl--th/9811028.

\bibitem{bub2} Y. Yu {\it et al.}, {\em Phys. Rev. Lett.} {\bf 84},
412 (2000).

\bibitem{casimir} M. Kardar and R. Golestanian, {\em Rev. Mod. Phys.}
{\bf 71}, 1233 (1999) and references therein.

\bibitem{strutin} V.M. Strutinsky and A.G. Magner,{\em Sov.  J.  Part.
Nucl.  Phys.} {\bf 7}, 138 (1976).

\bibitem{bohigas} O. Bohigas {\it et al.}, {\em Phys. Rep.}  {\bf
223}, 43 (1993); O. Bohigas {\it et al.}, {\em Nucl. Phys.} A {\bf
560}, 197 (1993); S. Tomsovic and D. Ullmo, {\em Phys. Rev.} E {\bf
50}, 145 (1994); S.D. Frischat and E. Doron, {\em Phys. Rev.} E {\bf
57}, 1421 (1998).

\bibitem{brack} M. Brack and R.K. Bhaduri, {\it Semiclassical
Physics}, Addison--Wesley, Reading, MA (1997); M. Brack, {\em
Rev. Mod. Phys.} {\bf 65}, 677 (1993) and references therein.

\bibitem{piotr} A. Bulgac and P. Magierski, astro--ph/0002377, {\em
Nucl. Phys.} A, in print; astro--ph/0007423, {\em Physica Scripta}, in
print; A. Bulgac. P. Magierski and A. Wirzba, unpublished.

\bibitem{pethick} C.J. Pethick and D.G. Ravenhall, {\em
Annu. Rev. Nucl. Part. Sci.}  {\bf 45}, 429 (1995) and references
therein.

\bibitem{rk} M.A. Ruderman, C. Kittel, {\em Phys. Rev.} {\bf 96}, 99
(1954).

\bibitem{heiselberg} H. Heiselberg {\it et al.}, {\em
Phys. Rev. Lett.}  {\bf 70}, 1355 (1992).

\bibitem{glen} M.B. Christiansen and N.K. Glendenning,
astro--ph/0008207.

\bibitem{madsen} G. Neergaard and J. Madsen, {\em Phys. Rev.} D {\bf
62}, 034005 (2000) and earlier references therein.

\end{thebibliography}
\end{document}